\documentclass[]{aa}  

\usepackage{graphicx}
\usepackage{amsmath}
\usepackage{mathtools}
\usepackage{caption, subcaption}
\usepackage{orcidlink}
\usepackage{txfonts}
\begin{document}

   \title{Modelling the connection between propagating disturbances and solar spicules}


   \author{S. J. Skirvin
          \inst{1,2}\orcidlink{0000-0002-3814-4232}
          \and
          T. Samanta\inst{3}\orcidlink{0000-0002-9667-6392}
          \and
          T. Van Doorsselaere\inst{1}\orcidlink{0000-0001-9628-4113}
          }

   \institute{Centre for mathematical Plasma Astrophysics, Department of Mathematics, KU Leuven, Celestijnenlaan 200B bus 2400, B-3001 Leuven, Belgium
   \and   Plasma Dynamics Group, Department of Automatic Control \& Systems Engineering, The University of Sheffield, UK
   \and   Indian Institute of Astrophysics, Koramangala, Bangalore, India \\
   \email{s.skirvin@sheffield.ac.uk}          
            }

   \date{Received --; accepted --}

 
  \abstract
   {}
   {Propagating (intensity) disturbances (PDs) are well reported in observations of coronal loops and polar plumes in addition to recent links with co-temporal spicule activity in the solar atmosphere. However, despite being reported in observations, they are yet to be studied in depth and understood from a modelling point of view.}
   {In this work, we present results from a 3D MHD numerical model featuring a stratified solar atmosphere which is perturbed by a p-mode wave driver at the photosphere, subsequently forming spicules described by the rebound shock model.}
   {Features with striking characteristics to those of detected PDs appear consistent with the co-temporal transition region dynamics and spicular activity resulting from nonlinear wave steepening and shock formation. Furthermore, the PDs can be interpreted as slow magnetoacoustic pulses propagating along the magnetic field, rather than high speed plasma upflows, carrying sufficient energy flux to at least partially heat the lower coronal plasma. Using forward modelling, we demonstrate the similarities between the PDs in the simulations and those reported in observations from IRIS and SDO/AIA.}
   {Our results suggest that, in the presented model, the dynamical movement of the transition region is a result of wave dynamics and shock formation in the lower solar atmosphere, and that PDs are launched co-temporally with the rising of the transition region, regardless of the wave-generating physical mechanisms occurring in the underlying lower solar atmosphere. However, it is clear that signatures of PDs appear much clearer when a photospheric wave driver is included. Finally, we present the importance of PDs in the context of providing a source for powering the (fast) solar wind.
}


   \maketitle
%

\section{Introduction} \label{sec:intro}

Intensity oscillations in the solar corona have been routinely observed over the last few decades. Propagating intensity (density) perturbations, otherwise known as propagating disturbances (PDs), reported in active region coronal loops have been interpreted as slow magnetoacoustic waves \citep{DeMoortel2002A, DeMoortel2002B, Wang2009_slowwaves, DeMoortel2012, Banerjee2021} and are widely thought to be connected to solar p-modes due to their observed periodicity in the $3$-$5$ minute regime. The connection between PDs and acoustic waves in sunspots is generally well understood, however, it is less clear whether waves or flows cause these PDs in plage and open field regions with conflicting interpretations from both observational \citep{Ban2009_slowwaves, tian2011, Tian2012, KrishnaPrasad2011, KrishnaPrasad2012, KrishnaPrasad2012_longperiods, Bryans2016} and modelling \citep{Verw2010, Wang2013, Demoortel2015} perspectives. In addition, slow magnetoacoustic waves have been well reported in coronal plumes \citep[e.g.][]{Ofman1999, Banerjee2000,mandal2018,Cho2021, Banerjee2021} and, although from observations it is unclear of the energy flux they carry, they may contribute to heating coronal plasma through compressive dissipation.

More recently, PDs detected in the solar atmosphere have been linked with jetting and solar spicule activity \citep{Jiao2015, Samanta2015, Pant2015, Jiao2016, Zhenyong2018,Bose2023}. In particular, \citet{Samanta2015} reported that the launching of the propagating intensity disturbance in AIA $171$ \AA{}/ $193$ \AA{} channels coincided temporally with the rising phase of the spicules seen in IRIS observations, confirmed upon comparison of time-distance diagrams. This led the authors to propose that magnetic reconnection generated both the waves and the spicules simultaneously, such that the waves can escape into the corona, whereas the dense and cool chromospheric material falls back to the surface due to gravity.

The impact of type II spicules on the mass and energy balance of the corona was studied by \citet{DePontieu2017}. The authors found that both shock waves and spicular flows contributed to the local plasma dynamics \citep[also see][]{Petralia2014}. \citet{DePontieu2017} discuss how the PDs may form `coronal loop strands' as a result of the inhomogeneous mass loading of individual magnetic field lines, which may have implications for the driving of the fast solar wind by providing an additional source of mass and energy to the solar corona \citep{Liu2015, Cranmer2017, Kumar2022,Chitta2023}.

It has been proposed that the chromosphere and transition region interface may behave as a leaky (acoustic) resonator, providing a source of propagating waves in the corona \citep{Botha2011,Snow2015,Felipe2019}. The acoustic resonator is a cavity formed between the steep temperature gradients provided by the photosphere and the transition region which produces distinct three minute oscillations in the chromosphere. However, there are also other wave generating mechanisms in the lower solar atmosphere which may provide a source of waves and intensity oscillations in the corona such as magnetic reconnection \citep{McLaughlin_et_al_2012}, nonlinear coupling of Alfv\'{e}n waves to other wave modes via the ponderomotive force \citep{Singh2022} and mode conversion from acoustic oscillations of the solar convection zone \citep{Schunker2006, Khomenko2012, Skirvin2024modeconv}

In this study, our aim is to provide evidence that PDs are inherently linked with the dynamical motion of the transition region and explain their link with spicular activity. This study is structured as follows: in Section \ref{sec:methods} we introduce the numerical model adopted in this study along with the relevant wave driver. In Section \ref{sec:results} we present the results of the simulations including the interpretation and observability of PDs and their connection with solar spicules. Finally, in Section \ref{sec:conclusions}, we provide a brief discussion of the implications of our results on providing an additional source of energy to the solar wind and the potential observability of PDs for next generation telescopes.

\section{Model} \label{sec:methods}

\begin{figure*}          
\includegraphics[width=0.99\textwidth]{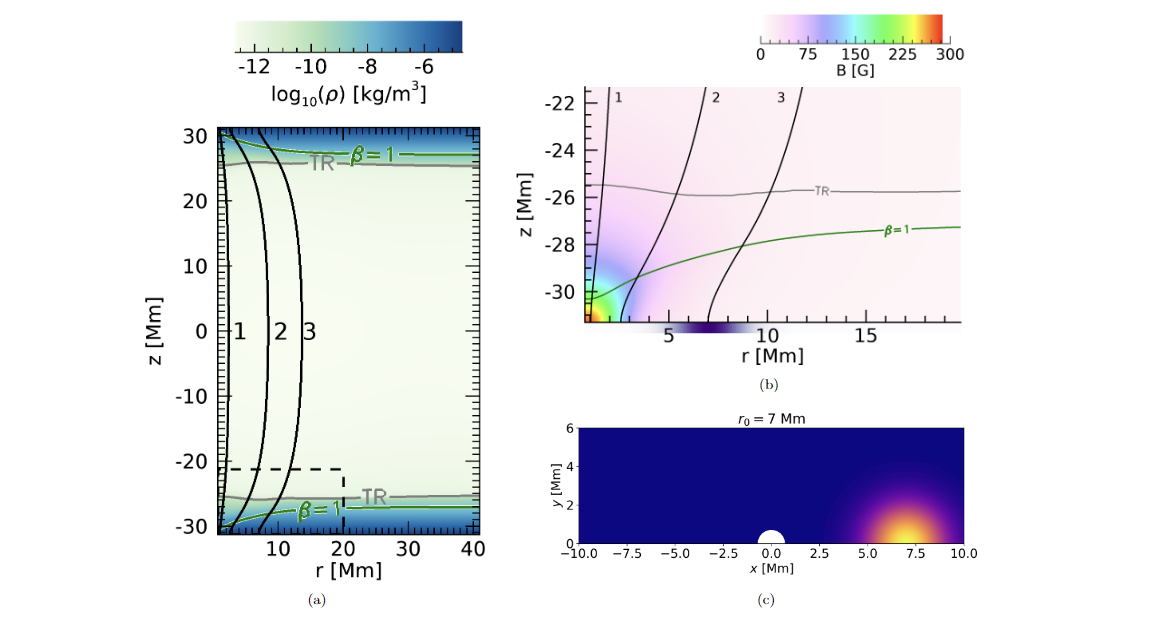}
\caption{(a) The initial background density (plotted on a log scale) in the full simulation domain at an azimuthal slice $\varphi=0$. The positions of field lines $1$, $2$ and $3$ indicated are used for discussion in the text. Contours for the plasma-$\beta = 1$ layer and transition region are shown by the green and grey lines, respectively. Panel (b) shows the initial magnetic field strength and topology outlined by the black dashed box in panel (a). The same configuration is found for all azimuthal angles due to the symmetry of the background equilibrium around the central axis. The location and strength of the (asymmetric) wave driver is highlighted by the purple bar on the horizontal axis. (c) The location of the wave driver employed with $r_0=7$ Mm (displayed on a Cartesian grid). The strength of the driver is shown in arbitrary units.
\label{fig:model}}
\end{figure*}

In order to study the connection between propagating disturbances and solar atmospheric dynamics, we conducted a 3D MHD numerical simulation using the PLUTO code \citep{Mign2007, Mign2018}. The background magnetohydrostatic model used in this work is adapted from the work of \citet{Serio1981} for closed coronal loop models and the initial total plasma density structuring in the domain is displayed in Figure \ref{fig:model}(a). The model adopts a cylindrical coordinate system ($r,\varphi,z$) and ranges from $0.73$ Mm to $41.01$ Mm in the radial direction, from $0$ to $\pi$ in the $\varphi$-direction, and from $-31.31$ Mm to $31.31$ Mm in the $z$-direction, with $192$ × $256$ × $768$ data points, respectively. The domain contains an enhancement of the magnetic field at the axis to represent a coronal loop spanning from photosphere to photosphere at the top and bottom boundaries of the domain \citep{Guar2014}. The magnetic field strength has a Gaussian profile in the radial direction, with magnitude of $273$ G on the axis which reduces to around $10$ G in the ambient atmosphere \citep{Reale2016} and is shown in Figure \ref{fig:model}(b). The background model is achieved through a numerical relaxation process, whereby a (vertical) magnetic field is inserted into the domain, and the system is allowed to evolve towards a numerical magnetohydrostatic equilibrium, such that the magnetic field expands in the atmosphere of the model. Even though the numerically relaxed atmosphere is taken as the equilibrium configuration for the simulations, there are still background velocities up to $18$ km s$^{-1}$ in the coronal volume \citep[e.g.][]{Reale2016, Riedl2021, Skirvin2023ApJ} as a result of total pressure gradients. However, the amplitudes of the background motions are considerably smaller than the characteristic speeds in the simulation (e.g. $<0.15 c_s$).

We introduce a wave driver at the bottom boundary of the simulation domain, in the photosphere, to resemble perturbations resulting from p-modes. The wave driver perturbs the vertical component of the velocity vector, in addition to perturbations of the plasma pressure and density and corresponds to the analytical solution of acoustic-gravity waves \citep{Mihalis1984,Riedl2021, Skirvin2023ApJ}. In this work, we present a single numerical simulation where we include a similar wave driver, with a Gaussian profile in strength in the radial direction and is centred on $r=7$ Mm \citep{Skirvin2024modeconv}. The driver is a 2D Gaussian in the $r-\varphi$ plane, displayed in Figure \ref{fig:model}(c), and is applied at the bottom boundary with a spatial width of $\sigma =2$ Mm, where $\sigma$ is the standard deviation of the localised Gaussian perturbation. 

As the background model is not in an equilibrium configuration, we also conduct a separate simulation excluding the wave driver, such that any wave properties can be isolated by subtracting the simulation without the driver from the simulation with the wave driver. However, we can also use the non-equilibrium state of the numerical background to investigate the response of the model to the dynamically evolving background solar atmosphere.

\section{Results} \label{sec:results}
\subsection{Wave evolution}\label{subsec:simulations}

Let us consider a scenario whereby the simulation domain is perturbed by a localised photospheric wave driver \citep[see e.g.][]{Riedl2021, Skirvin2023ApJ, Gao2023}, indicated in Figure \ref{fig:model} and outlined in Section \ref{sec:methods}. In this case, the waves driven at the bottom boundary of the simulation steepen into shocks as they propagate up through the stratified solar atmosphere. The shocks repeatedly interact with the transition region and cause it to lift upwards, as described by the rebound shock model of spicule formation \citep{holl1982, sue1982}. However, in a previous study using the current numerical model, \citet{Riedl2021} suggested that standing waves formed by the interaction of the driven waves with the background motions are responsible for the oscillation of the TR height. It is possible that the formation of standing waves may play a role in the TR dynamics and, in addition, that the oscillation of the TR interacting with the driven waves may form the standing waves observed in the chromosphere in the simulations. 
\begin{figure*}
\begin{subfigure}{0.33\textwidth}
    \includegraphics[width=\textwidth]{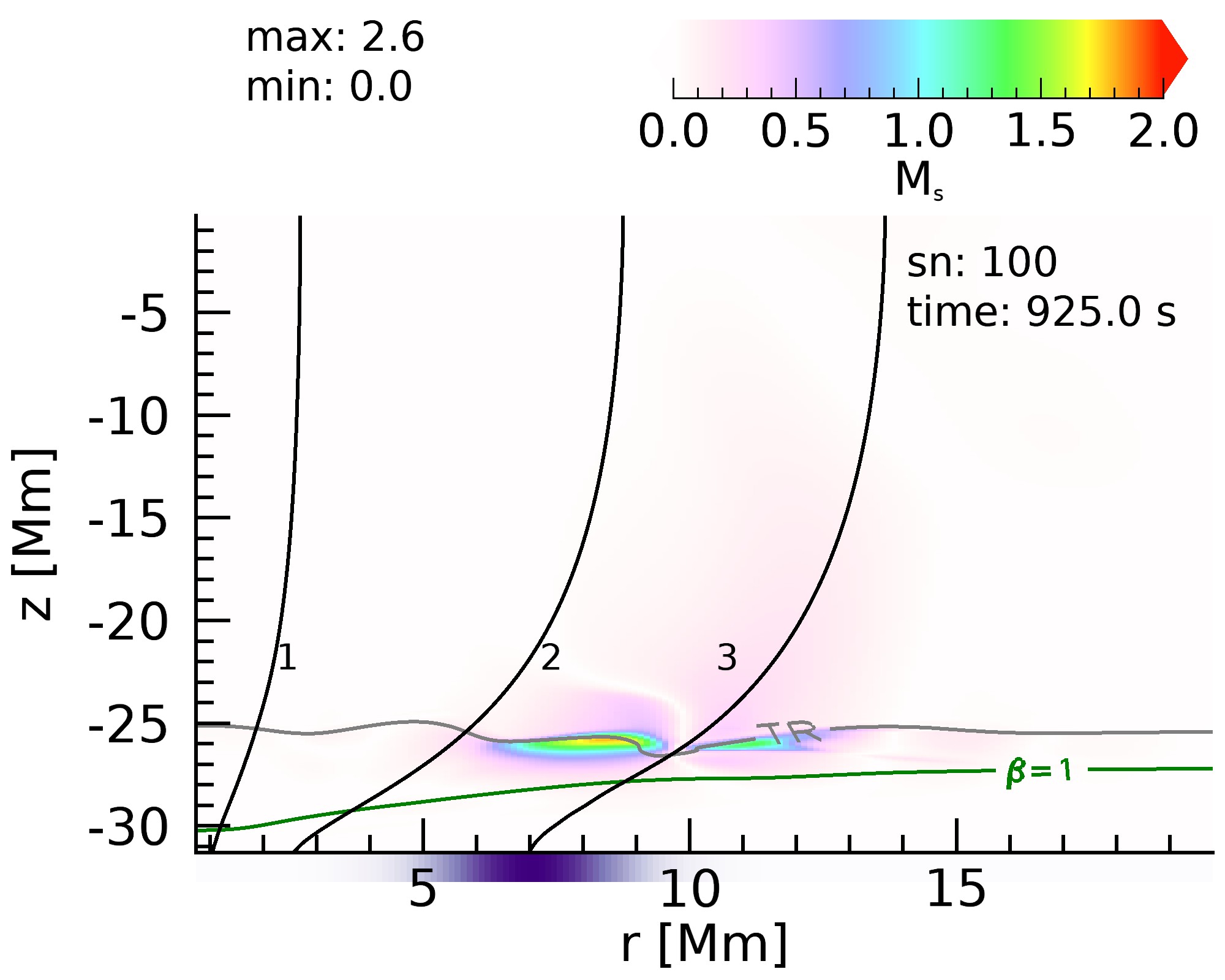}
    \caption{}
\end{subfigure}
\begin{subfigure}{0.33\textwidth}
    \includegraphics[width=\textwidth]{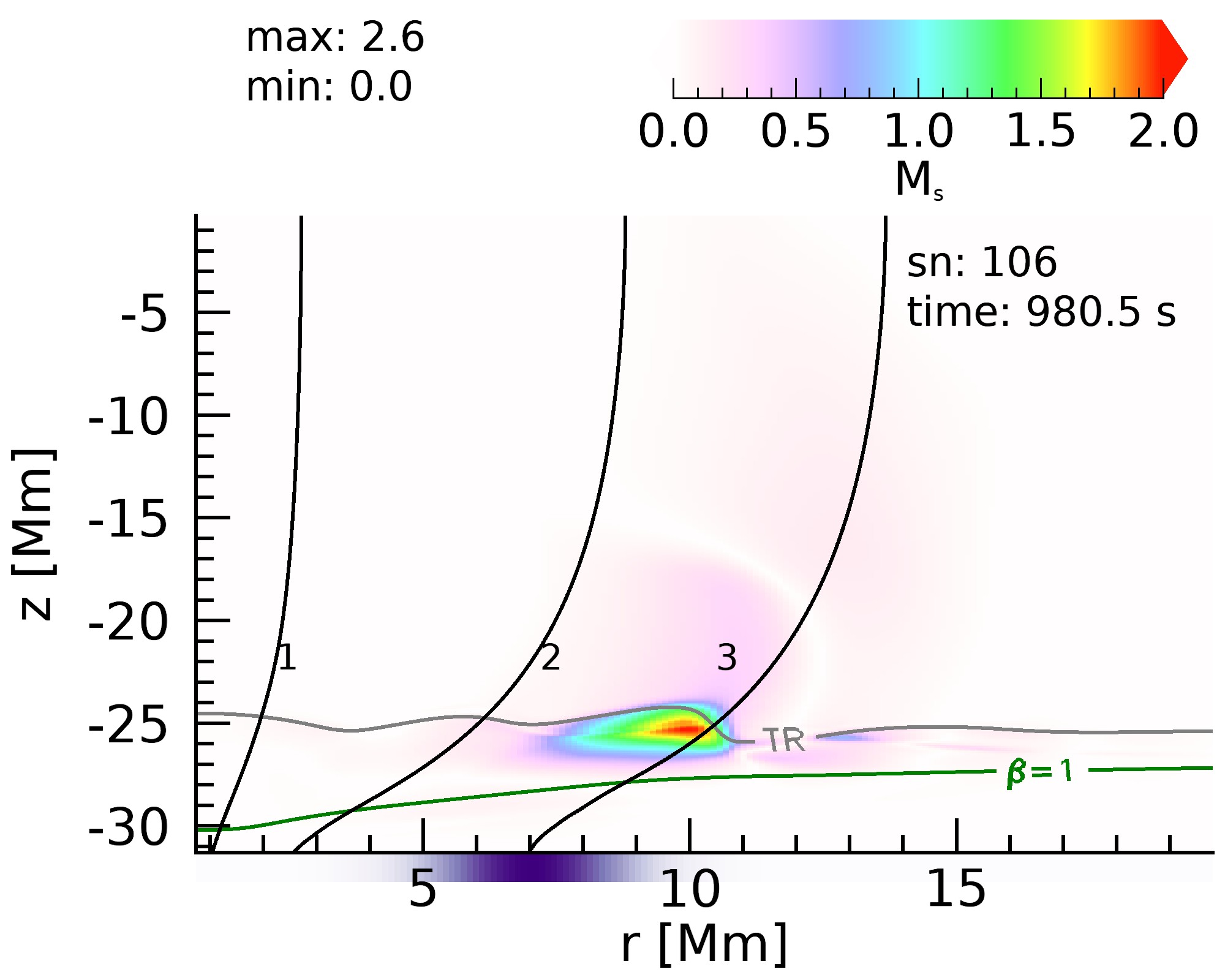}
    \caption{}
\end{subfigure}
\begin{subfigure}{0.33\textwidth}
    \includegraphics[width=\textwidth]{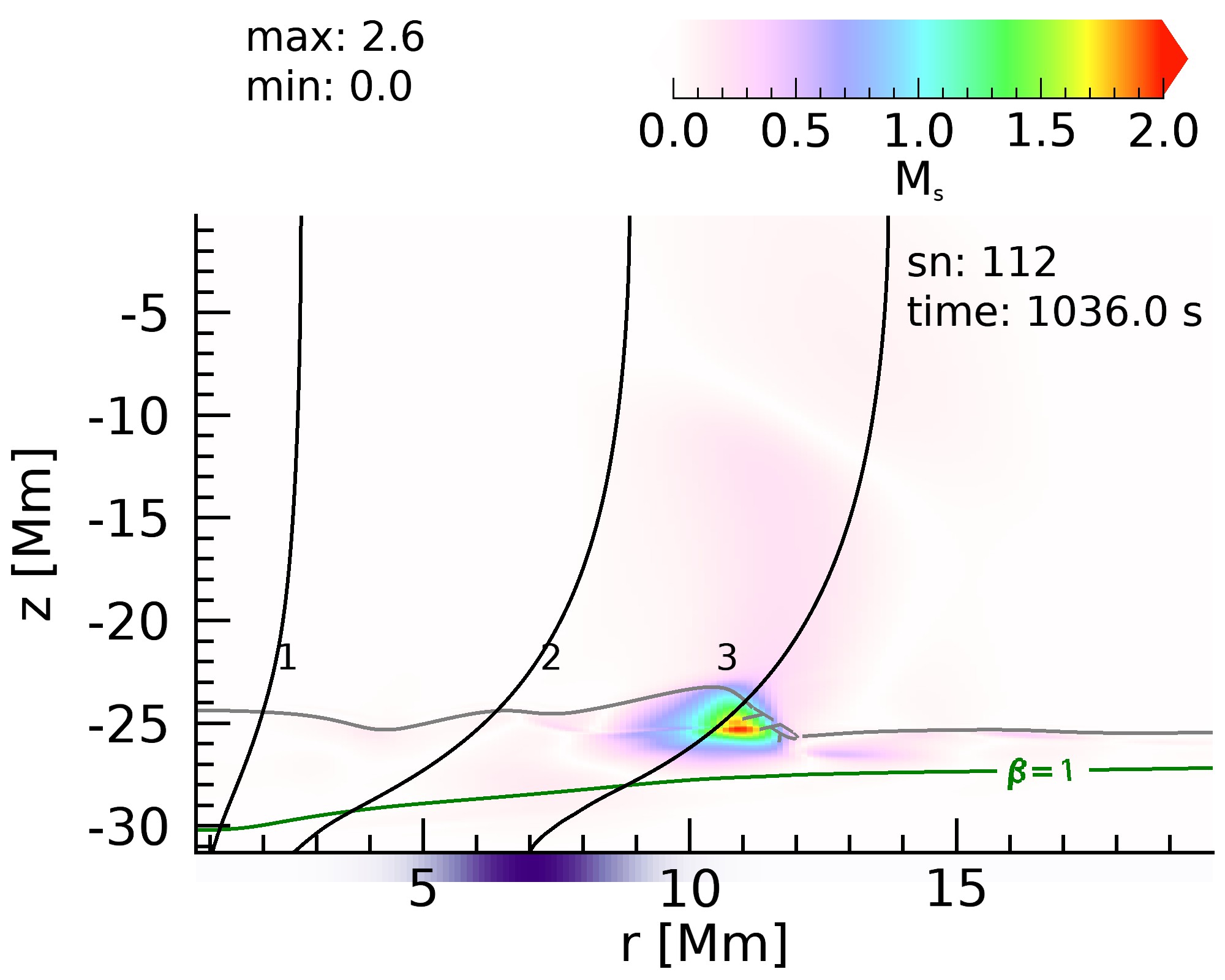}    
    \caption{}
\end{subfigure}
\caption{Snapshots of the parallel Mach number $M_s = |\hat{v}_{\parallel}|/c_s$ in the wave-driven simulation at times corresponding to (a) $t=925$~s, (b) $t=981$~s and (c) $t=1036$ s. An animated version of this figure is available online in addition to an animated version of the Mach values in the un-driven simulation.
\label{fig:mach_vals}}
\end{figure*}
The shock formation is highlighted in Figure \ref{fig:mach_vals} where the Mach number of velocity parallel to the magnetic field is displayed in the simulation with a wave driver. The parallel Mach number is calculated as $M_s = |\hat{v}_{\parallel}|/c_s$, where $c_s$ is the local sound speed. It is evident that shocks are formed below the transition region along the field line where the wave driver is rooted. The nonlinear evolution of the wave amplitude generates supersonic motions with a maximum Mach value of $M_s = 2.6$. The propagating disturbance is also visible in Figure \ref{fig:mach_vals} in the coronal domain, although the Mach value associated with the PD is subsonic with $M_s = 0.5$. A similar situation is encountered in the simulation without a wave driver, however in this case, the maximum Mach values are not supersonic in nature, as such may not be defined as shocks, with maximum value $M_s=0.8$, although it is clear that even in the un-driven case, the background motions can become highly nonlinear due to the steep density gradients in the chromosphere.
\begin{figure}
    \centering    \includegraphics[width=0.49\textwidth]{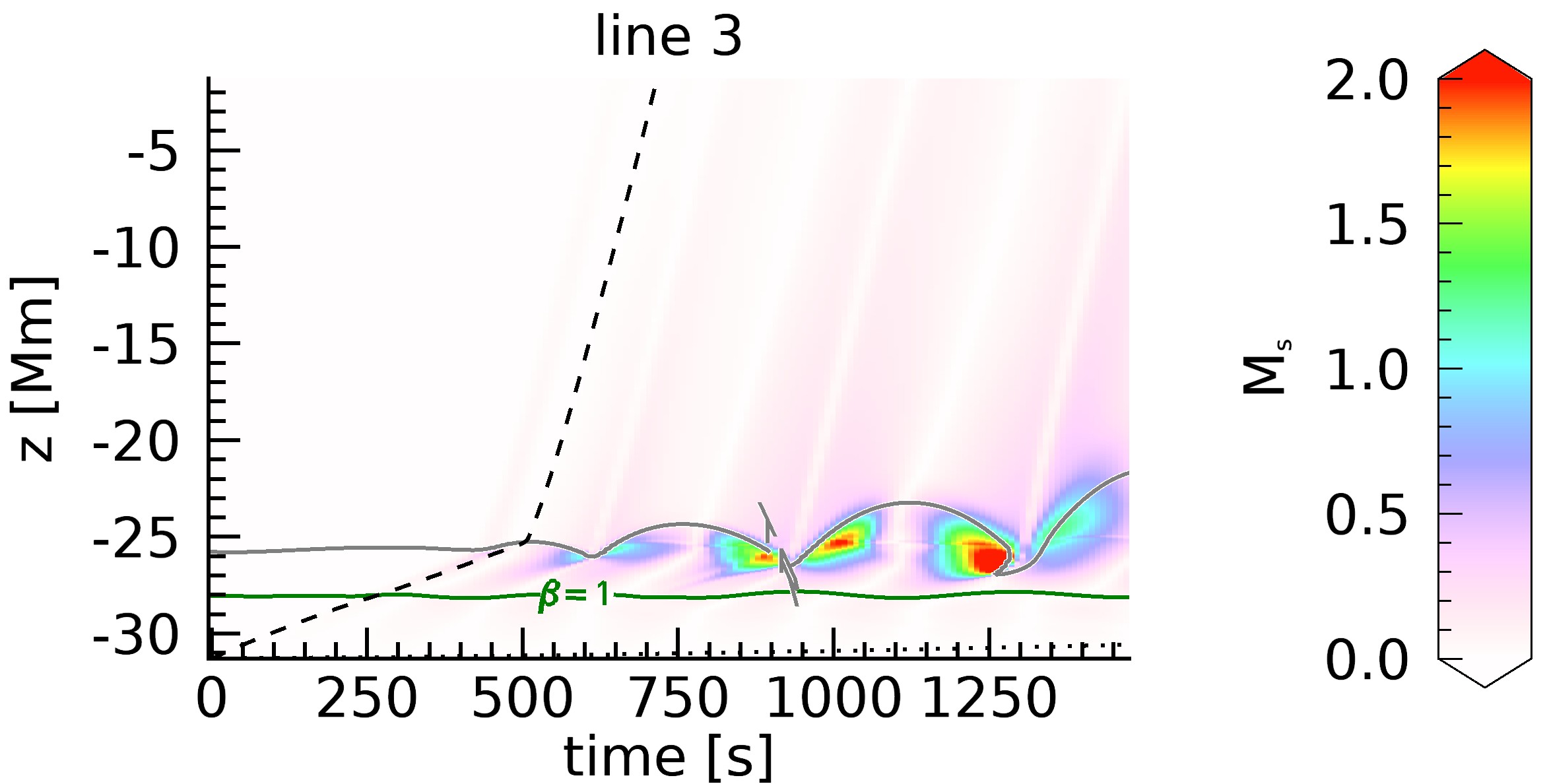}
    \caption{Time-distance plot along field line $3$ in the wave-driven simulation displaying the parallel Mach number $M_s = |\hat{v}_{\parallel}|/c_s$.}
    \label{fig:mach_td}
\end{figure}
Figure \ref{fig:mach_td} displays the time-distance diagram of the parallel Mach number along field line $3$ for the wave-driven simulation. It is evident that there is a strong correlation between the formation of shocks, denoted by supersonic Mach values, below the transition region and the launching of the transition region. Moreover, the launching of the PDs can also be seen in the time-distance diagram in Figure \ref{fig:mach_td} as the coronal enhancements of the Mach value $M_s = 0.6$ as the shock continues to propagate into the corona.

\begin{figure}
    \centering    \includegraphics[width=0.49\textwidth]{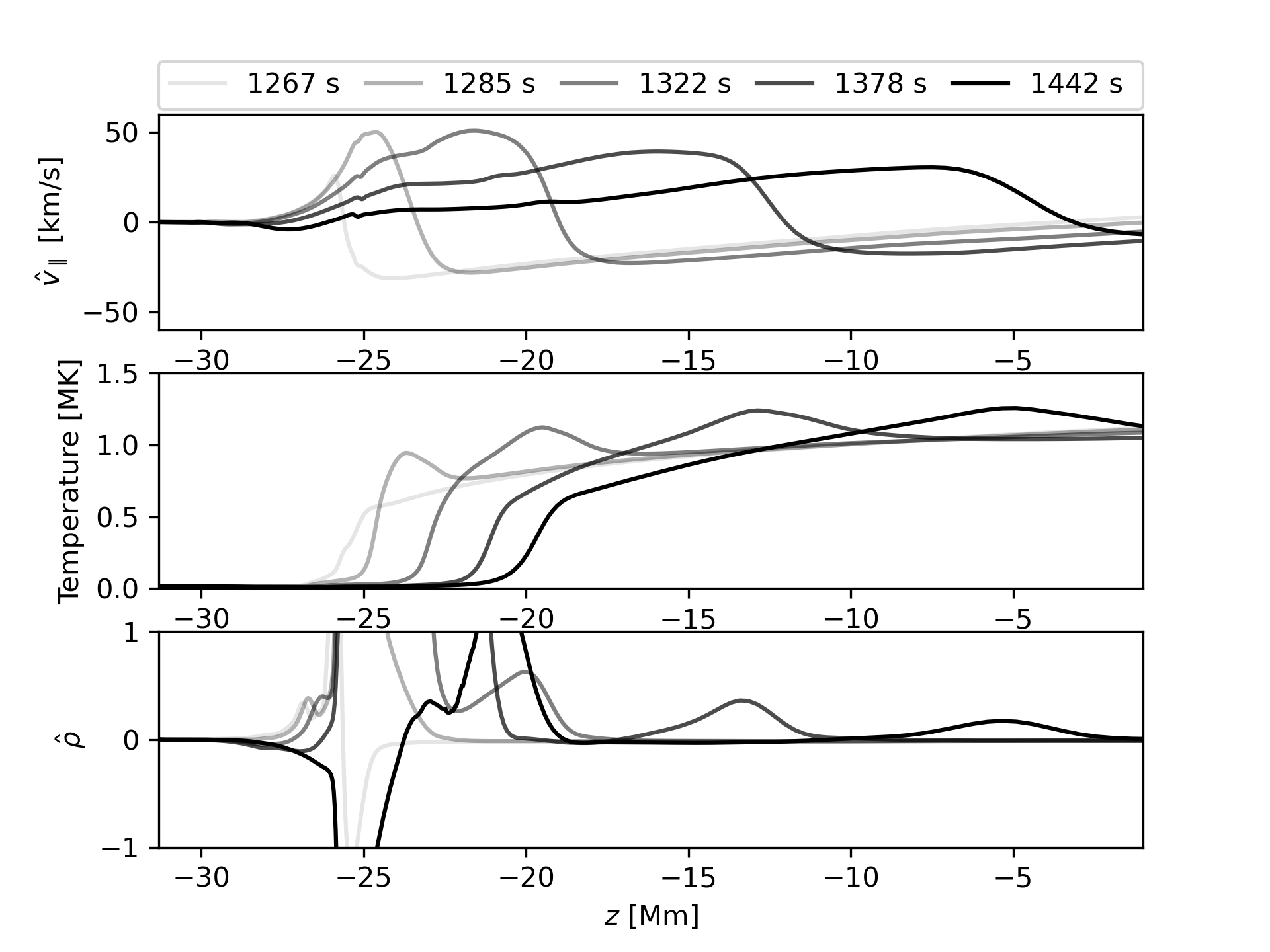}
    \caption{The time evolution of the velocity perturbation along field line 3 ($\hat{v}_{\parallel}$) in the simulation with a photospheric wave driver, the temperature along field line 3 and the density perturbation ($\hat{\rho}$) along field line 3 calculated using Equation (\ref{density_pert}). An animated version of this figure is available online.}
    \label{fig:time_evolution}
\end{figure}
The time evolution of the shock formation is displayed in Figure \ref{fig:time_evolution}. We calculate the evolution of the density perturbation as:
\begin{equation}\label{density_pert}
    \hat{\rho}(t,z) = \frac{\rho_t(z) - \rho_{t-1}(z)}{\rho_0(z)},
\end{equation}
where $t$ denotes time and $\rho_0$ indicates the initial value of the density at that height ($z$) in the domain. The velocity perturbation along the magnetic field $\hat{v}_{\parallel}$ is shown in the top panel of Figure \ref{fig:time_evolution} and it is evident that as the slow magnetoacoustic wave propagates up in the stratified lower solar atmosphere, its amplitude non-linearly steepens characteristic of a shock formation. The shock then hits the transition region and causes it to lift upwards due to the strong pressure force generated from the shock, as can be seen in the temperature evolution in Figure \ref{fig:time_evolution}. For example, at $t=1267$ s, the transition region is located around $z=-26$ Mm in the simulation domain, however, at a later time of $t=1442$ s, the transition region is lifted to $z=-20$ Mm, in other words, the transition region is lifted by roughly $6$ Mm and this would be observed as a spicule.
\begin{figure}
    \centering    \includegraphics[width=0.49\textwidth]{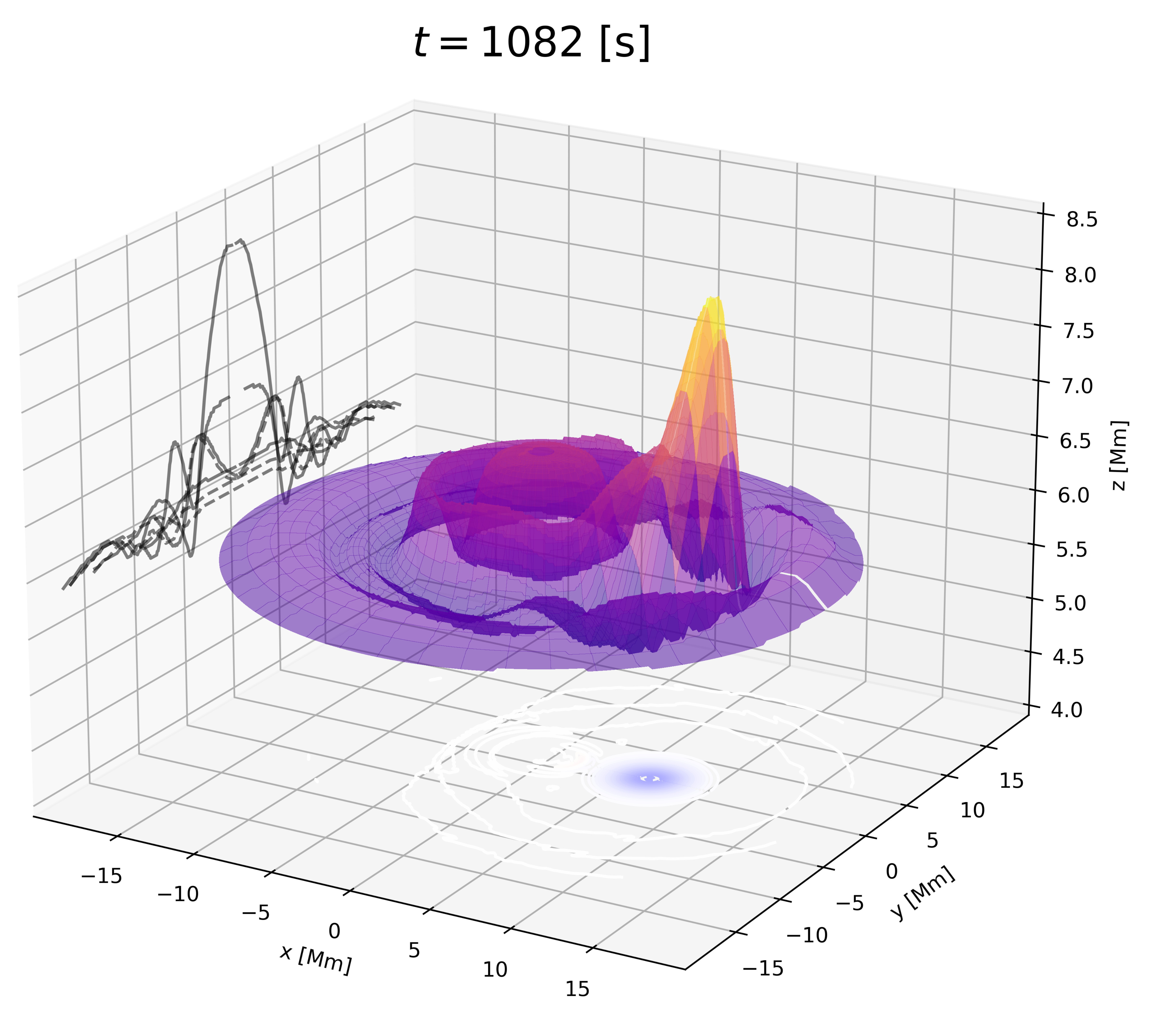}
    \caption{A 3D representation of the height (above the bottom boundary of the simulation) of the transition region, calculated as a temperature contour of $40,000$ K at a specific snapshot in time of $t=1082$ s. The strength of the $v_z$ perturbation is shown on the bottom boundary plane and the shadow contour of the transition region height is projected against the $y-z$ plane to highlight the surface waves formed at the transition region resulting from previous spicule launches.}
    \label{fig:3D_spicule}
\end{figure}
A 3D representation of the spicule formation is highlighted in Figure \ref{fig:3D_spicule} at a snapshot in time at $t=1082$ s. This figure displays the height of the transition region above the bottom boundary of the simulation and produces a thin jet-like feature resembling a spicule which possesses heights consistent with spicule observations \citep{Sterling_2000, tsi2012, Skirvin2023_JET}. The 3D contour is shown for a temperature of $40,000$~K. In figure \ref{fig:3D_spicule}, the position of the wave driver is shown by the contour on the bottom boundary plane. Moreover, we project the shadow of the transition region against a vertical plane to highlight the surface waves generated at the transition region which are formed from spicules at earlier times and are representation of so called `transition region quakes' \citep{scu2011}. The spicule can also be seen in the bottom panel of Figure \ref{fig:time_evolution} as the large density perturbation between $z=-26$ Mm and $z=-20$ Mm.  Moreover, there is an additional perturbation, seen in the temperature and $\hat{\rho}$ plots, which propagates higher into the corona at the same speed as the front of the $\hat{v}_{\parallel}$ perturbation. This feature propagates at the local sound speed (see e.g. Figure \ref{fig:0deg_shifted_rho_t-d}) and represents the propagating disturbance.

\subsection{Propagating density disturbances}

\begin{figure*}
\begin{subfigure}{0.5\textwidth}
    \includegraphics[width=\textwidth]{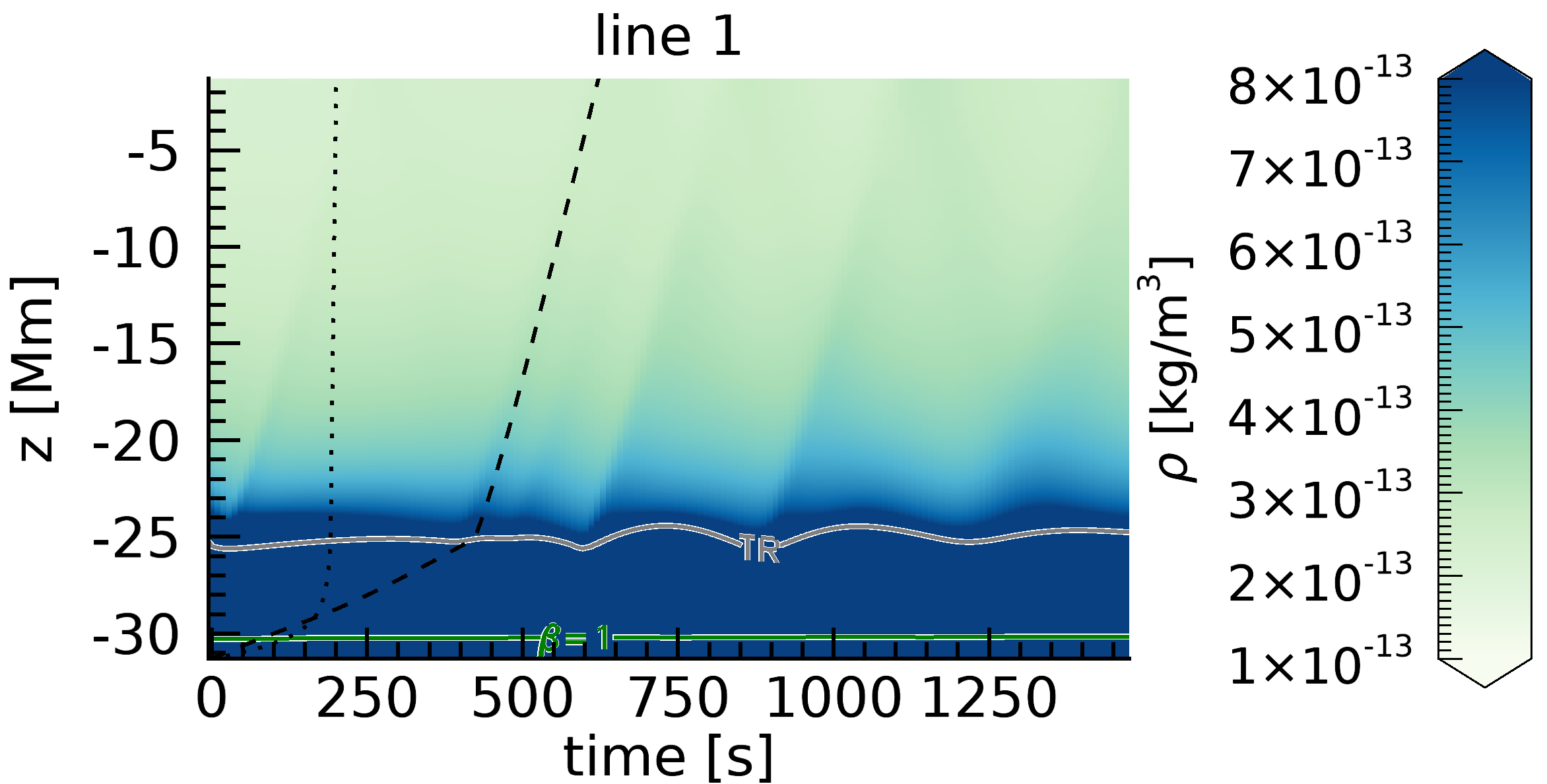}
    \caption{}
\end{subfigure}
\begin{subfigure}{0.5\textwidth}
    \includegraphics[width=\textwidth]{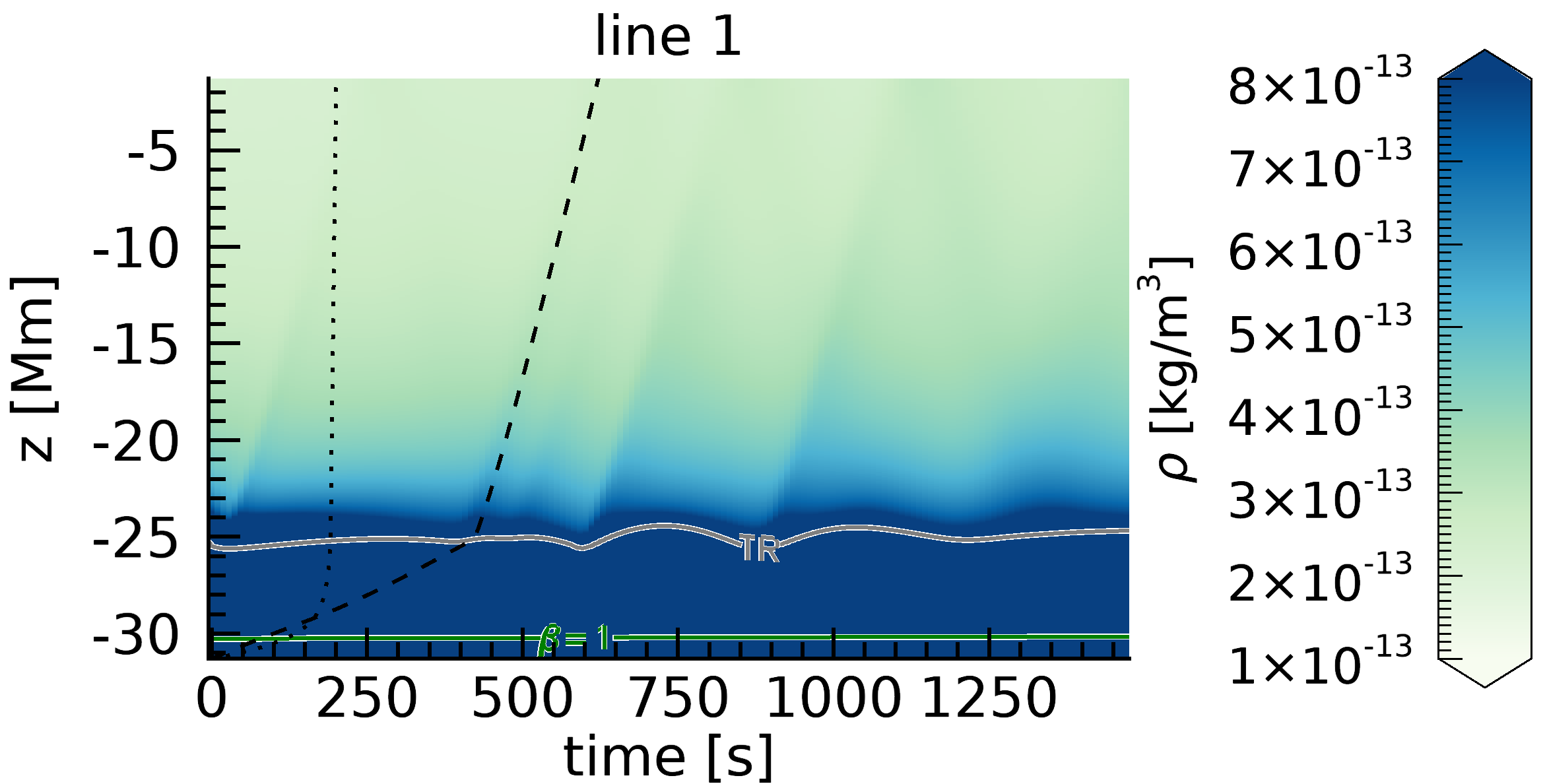}
    \caption{}
\end{subfigure}
\begin{subfigure}{0.5\textwidth}
    \includegraphics[width=\textwidth]{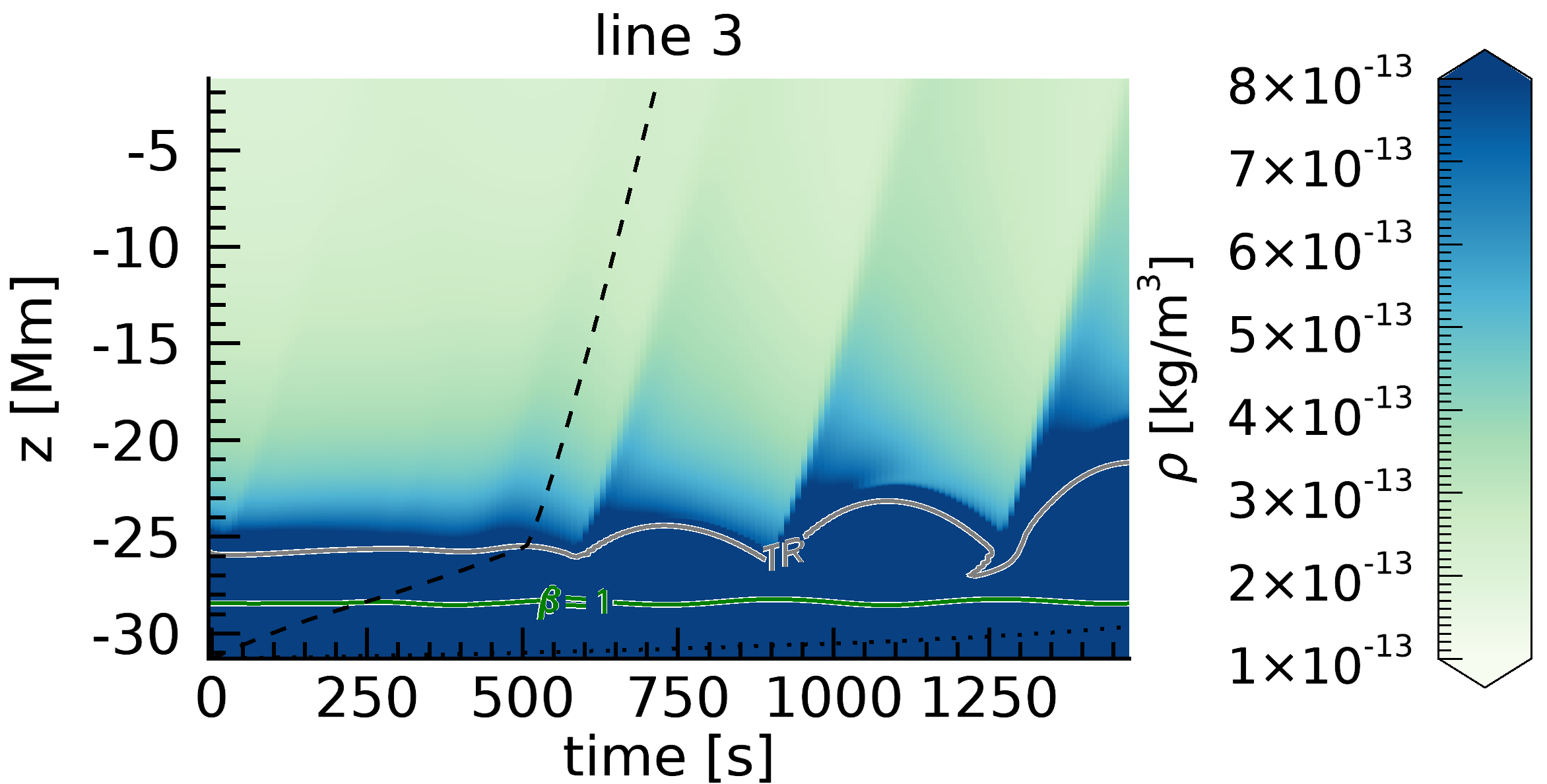}    
    \caption{}
\end{subfigure}
\begin{subfigure}{0.5\textwidth}
    \includegraphics[width=\textwidth]{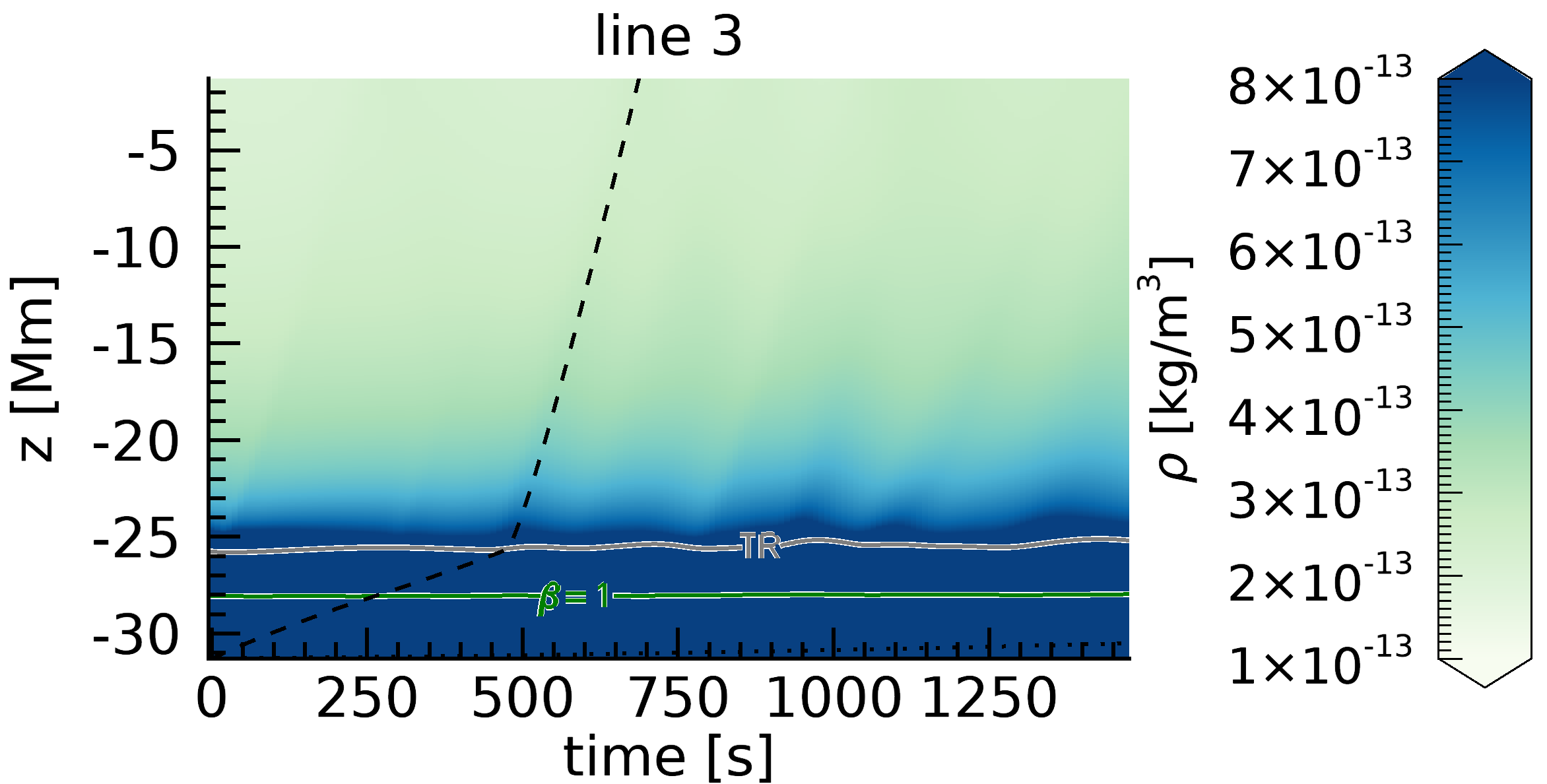}
    \caption{}
\end{subfigure}
\caption{Time-distance plots of the density in the simulation where an acoustic gravity wave driver is positioned at $r = 7$ Mm, for the three field lines indicated in Figure \ref{fig:model} at an azimuthal slice $\varphi=0$. Locations of the transition region and plasma-$\beta = 1$ layer are indicated by the grey and green contours respectively. We also display the local sound speed (dashed line) and local Alfv\'{e}n speed (dotted line) in each plot. The colour bar is saturated at the upper limit to aid visualisation.
\label{fig:0deg_shifted_rho_t-d}}
\end{figure*}

Seen in the time-distance diagrams in Figure \ref{fig:0deg_shifted_rho_t-d} (left hand panels), the transition region (grey contour), defined as a temperature contour corresponding to $T=40,000$ K, displays a parabolic trajectory characteristic of type I spicules which is also evident in the time-distance plots presented in \citet{Samanta2015}. The parabolic trajectory is more noticeable for the time-distance plots made on field line 3 (panel e), on which the localised wave driver is rooted. Interestingly, we observe density enhancements propagating at the local sound speed, which begin at every `trough' of the transition region in the time-distance plots. These density enhancements are temporally associated with the rising phase of the transition region in the simulation and can be seen propagating several megameters into the corona. The magnitude of the propagating disturbance intensity increases as the simulation evolves, as the driven shocks interact with a transition region whose oscillating amplitude is increasing, ejecting more chromospheric material higher into the corona. In other words, the intensity of the propagating disturbances increases with increasing total momentum between the wave-transition region interaction. Of course, this exact behaviour is unlikely to occur in nature as there are no monochromatic wave drivers which are fixed in an precise location on the solar surface. The periodicity of the PDs matches that of the periodicity of the wave driver ($370$ s) as they are simply a result of the response to the transition region interacting with the driven compressible waves. This result is consistent with recent reports of a periodic brightening phenomenon as a response to solar jets modulated by p-modes \citep{Cai2024}. Similar features representing propagating disturbances are also observed on field lines $1$ and $2$, positioned further away from the localised wave driver, however, rooted in regions of stronger magnetic field (see Figure \ref{fig:model}). This suggests that PDs may not be solely a back-reaction from wave driving at the photosphere, although the propagating horizontal wave fronts as a result of the driver may interfere with the transition region in these locations.

In order to verify the idea that the propagating disturbances are not solely a result of the driven waves interacting with the transition region, we conducted the same analysis as before, however, for the simulation with no photospheric wave driver. We remind the reader that the background atmosphere is not in a complete magnetohydrostatic equilibrium, therefore, there are background motions which we can exploit to study the relationship between propagating disturbances and the dynamically evolving solar atmosphere. Shown in Figure \ref{fig:0deg_shifted_rho_t-d} (right panels) are the resulting time-distance plots for the total density on the three selected field lines in the simulation without a wave driver. Distinct density enhancements can be seen along all three selected field lines, which propagate at the local sound speed. The intensity of the PDs is greater on field line 1 compared to the other field lines as field line 1 is rooted in a region of stronger magnetic field, associated with greater background velocities which subsequently cause a greater displacement of the transition region. This is a result of the magnetic field lines being more vertical when rooted in a region of stronger magnetic field, therefore, waves travelling along these field lines experience stronger vertical stratification and form stronger shocks when compared to motions propagating along inclined field lines. As a result, the amplitudes of the generated slow magnetoacoustic waves are stronger on field line 1 compared to field line 3, which is highlighted by the increased density enhancement in the time-distance diagrams. Similar to the wave driven simulation, the PDs appear to emanate from the rising phase of the transition region, which is consistent with observations. Moreover, the periodicity of the PDs in the simulation without a photospheric wave driver is reminiscent of the three minute periodicity, especially on field line 1 highlighted in Figure \ref{fig:0deg_shifted_rho_t-d}(b). This is consistent with the chromospheric acoustic resonator idea, indicating that the presence of PDs in the corona may originate from different physical mechanisms.

\subsection{Acoustic wave energy flux associated with PDs}

If the interpretation of PDs as slow magnetoacoustic waves is correct, then we should expect these disturbances to carry (acoustic) wave energy flux along the magnetic field. To investigate this, we plot the time-distance diagrams for the hydrodynamic (HD) wave energy flux parallel to the magnetic field, $F_{\text{HD}, \parallel}$, given by the formula:
\begin{equation}\label{eqn_flux}
    F_{\text{HD}, \parallel} = \left( \frac{\rho v^2}{2} + \rho \Phi + \frac{\gamma}{\gamma+1}p \right)v_{\parallel},
\end{equation}
where $\rho$ is the plasma density, $\Phi$ denotes the gravitational potential, $\gamma$ is the adiabatic index, $p$ the plasma pressure and $v_{\parallel}$ the velocity parallel to the magnetic field. In the case of the simulation without the wave driver, we plot the total full HD flux whereas for the simulation with the wave driver, we calculate the perturbed wave energy flux by subtracting the simulation without the wave driver from the simulation with the wave driver and inserting the perturbed quantities into Equation (\ref{eqn_flux}).
\begin{figure}
\begin{subfigure}{0.45\textwidth}
    \includegraphics[width=\textwidth]{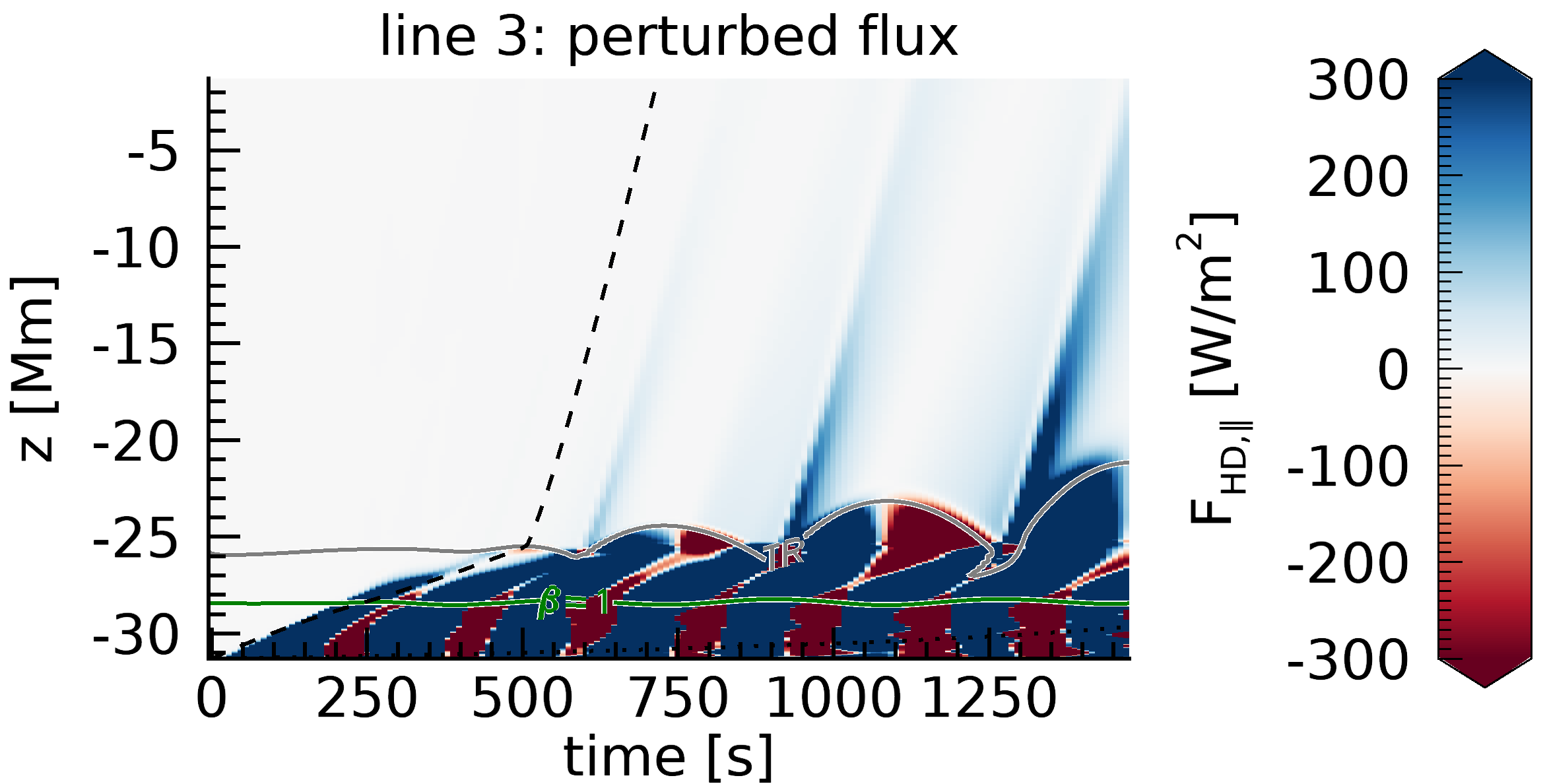}
    \caption{}
\end{subfigure}
\begin{subfigure}{0.45\textwidth}
    \includegraphics[width=\textwidth]{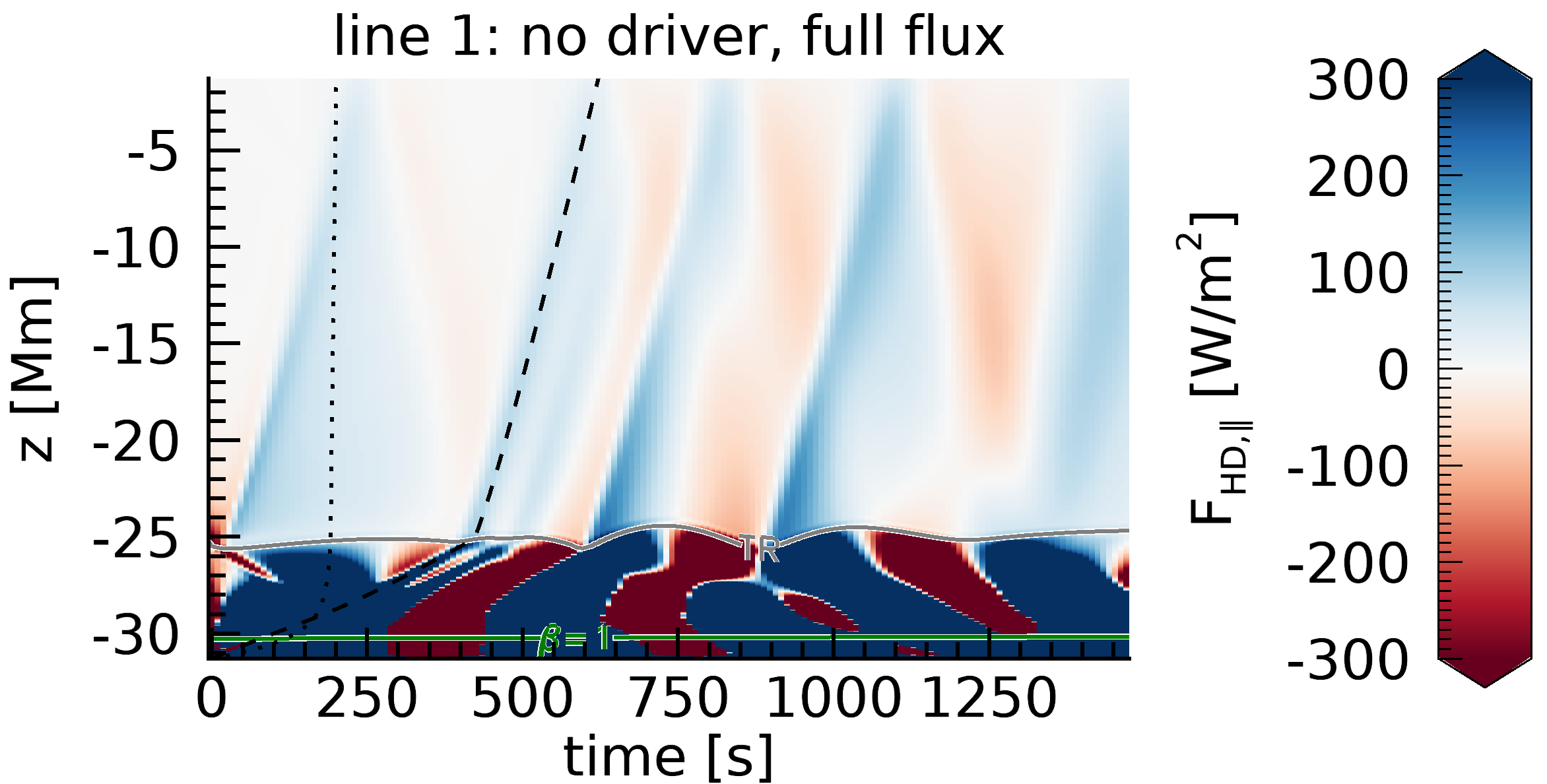}
    \caption{}
\end{subfigure}
\caption{Time-distance diagrams of (a) the perturbed hydrodynamic wave energy flux parallel to the magnetic field along field line 3 at an azimuthal slice $\varphi=0$ for the driven simulation and (b) the total hydrodynamic wave energy flux along field line 1 for the un-driven simulation. In both plots the transition region (T=$40,000$ K) is denoted by the grey contour whereas the green contour outlines the $\beta=1$ layer. The local sound speed (dashed line) and local Alfv\'{e}n speed (dotted line) are also plotted on each field line.}
\label{fig:hd_flux_tds}
\end{figure}

The resulting time-distance plots of HD flux parallel to the magnetic field are displayed in Figure \ref{fig:hd_flux_tds}. We can immediately see a clear relationship between the flux emitted into the corona and the density perturbations displayed in Figure \ref{fig:0deg_shifted_rho_t-d}. At times corresponding to the transition region minima, there is HD flux emitted into the corona, which has previously been reported by \citet{Riedl2021}, however no such link was made with signatures of propagating disturbances at the time. The HD flux parallel to the magnetic field clearly propagates at the local sound speed, even for the case when no wave driver is applied. As the small background motions (as a result from the non-equilibrium background state) only have a maximum amplitude of around $18$ km s$^{-1}$ in the corona, and the local sound speed in this region is roughly $100$ km s$^{-1}$, then these PDs cannot be a result of mass flows, therefore they must be attributed to slow magnetoacoustic waves. Moreover, if these PDs were signatures of mass flows, we would expect them to follow a parabolic trajectory in the time-distance plots, similar to the parabolic trajectory exhibited by type I solar spicules, whereas it is evident that the HD flux propagates along the magnetic field with no obvious deceleration. The agreement between the intensity/density enhancements and the HD flux emission solidifies the link with PDs in the simulation and confirms their interpretation as slow magnetoacoustic waves. The HD flux parallel to the magnetic field can be seen to decrease with height in the solar atmosphere, this may be due to damping via thermal conduction which is included in the numerical model. In addition, the HD flux carried by the PDs can be of the order of $>200$ Wm$^{-2}$, which is thought to be sufficient to balance the energy losses in the quiet Sun \citep{with77, Klimchuk2006}, suggesting the possibility that PDs may play a role in the energy budget of the solar atmosphere through energy dissipation mechanisms such as compressive viscosity, which often possesses nonphysical magnitudes in magneto-convection simulations \citep{Rempel2017}.

\subsection{Mass flux associated with simulated PDs}

We estimate the mass flux associated with PDs as:
\begin{equation}\label{mass_flux_eqn}
    \text{Mass flux} = 4\pi r^2 \rho v f_s,
\end{equation}
where $r=1.0$\(R_\odot\), $\rho$ is the plasma density associated with PDs in connection with solar spicules, $v$ is the propagation speed of PDs (taken to be the local sound speed in the corona) and $f_s$ is a filling factor related to the area of the solar chromosphere filled with spicules, taken to be $0.015 = 1.5\%$. 
\begin{figure}
 \centering   \includegraphics[width=0.45\textwidth]{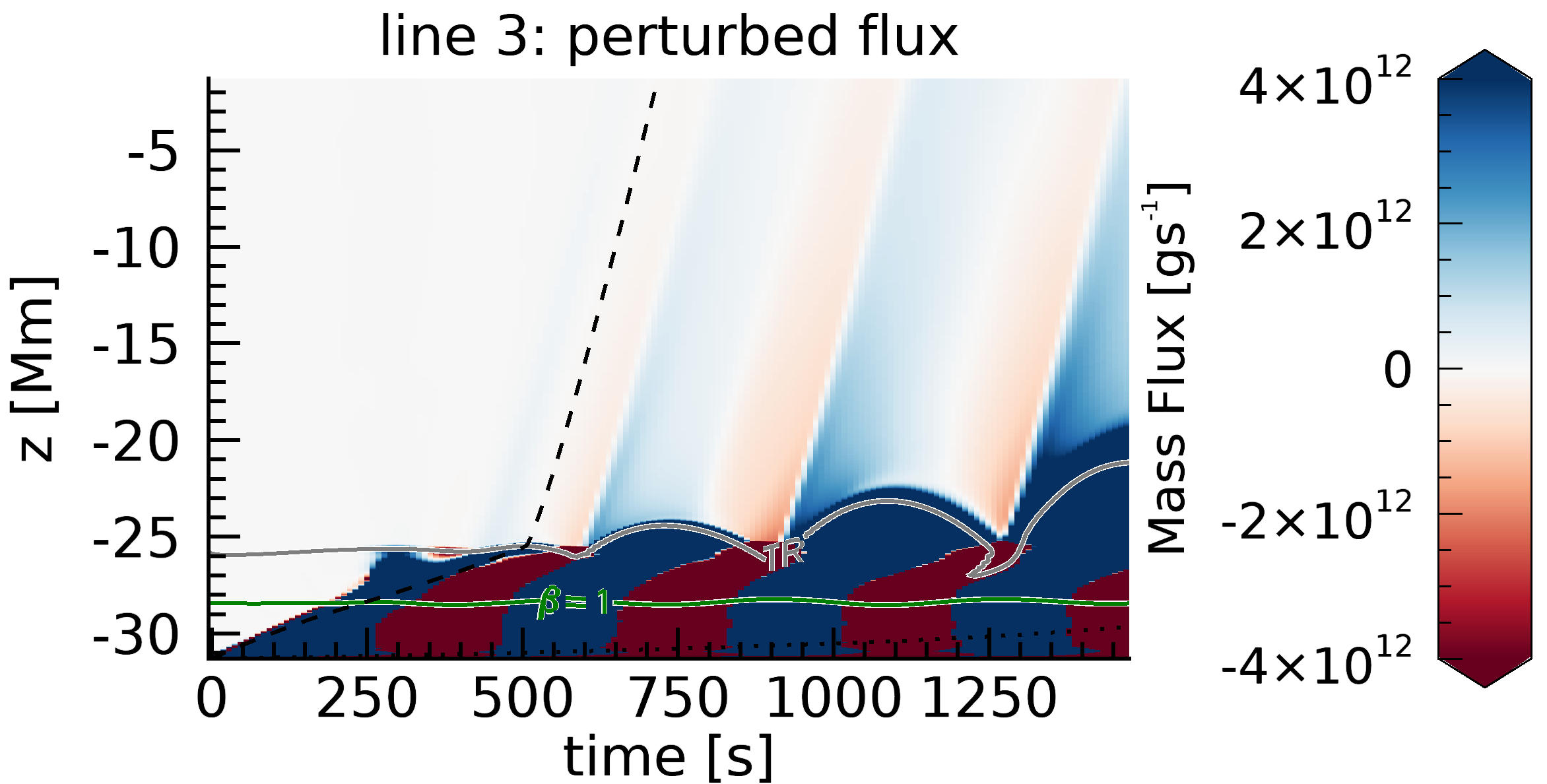}
    \caption{Time-distance plot of the perturbed mass flux associated with PDs from the simulation with a wave driver along field line 3 at an azimuthal slice $\varphi=0$ calculated using Equation (\ref{mass_flux_eqn}) with the density perturbation $\hat{\rho}$ determined from the simulation with a wave driver minus the simulation without a wave driver.}
\label{fig:massflux_tds}
\end{figure}
The time-distance diagrams of the calculated mass flux of PDs associated with the simulation with the photospheric wave driver is shown in Figure \ref{fig:massflux_tds}. There is clearly a strong relationship between the mass flux emitted into the solar corona with the launching of the transition region. This result agrees with the co-temporal relationship between the rising phase of spicules and the generation of a propagating disturbance. The top panel in Figure \ref{fig:massflux_tds} displays the mass flux calculated using Equation (\ref{mass_flux_eqn}) by inputting the full density in the simulation, whereas Figure \ref{fig:massflux_tds}(b) shows the perturbed mass flux whereby the perturbed value of $\hat{\rho}$ is used in the calculation. Regardless of the variable $\rho$ or $\hat{\rho}$ used in Equation (\ref{mass_flux_eqn}), the computed mass flux associated with PDs possesses the same order of magnitude. 

If we were to take single values consistent with those from the simulations, we can obtain a conservative rough estimate of the perturbed mass flux of PDs as $1 \times 10^{12}$ g s$^{-1}$, see Figure \ref{fig:massflux_tds}(b), which is consistent with that reported from small-scale `jetlets' at the base of solar plumes associated with magnetic reconnection \citep{Kumar2022} and larger than supersonic density fluctuations reported by \citet{Cho2020}. This value corresponds to a mass-loss resulting from PDs due to solar spicules as $1.57 \times 10^{-14} \ M_\odot$ yr $^{-1}$ which corresponds to $79$ \% of the global solar wind \citep{Cohen2011}. Therefore, PDs likely play an important role in supplying mass to the solar wind and given the ubiquity of solar spicules throughout the entire solar cycle, they offer a consistent source of energy to accelerate the solar wind.

However, the mass flux transported by PDs depends on the strength of the initial slow magnetoacoustic shock, responsible for launching the transition region. Therefore, the mass flux of PDs is likely to fluctuate around an average of the possible mass densities associated with solar spicules which will influence the contribution of PDs to the mass-loss rate of the solar wind.

\subsection{Forward Modelling}\label{subsec:forwardmodelling}

In observations such as those reported by \citet{Samanta2015} and \citet{Jiao2015}, PDs appear as clear intensity enhancements in the AIA 171 \AA{} channel. So far, we have provided evidence of PDs in the simulation using the plasma density as a diagnostic, however, in order to provide an accurate comparison with observations, it would be beneficial to convert the numerical output to observational quantities. Therefore, we apply the FoMo code \citep{TomVD2016} to forward model the simulation output into observable features, as would be seen using AIA 171~\AA. 
\begin{figure*}          
\includegraphics[width=0.98\textwidth]{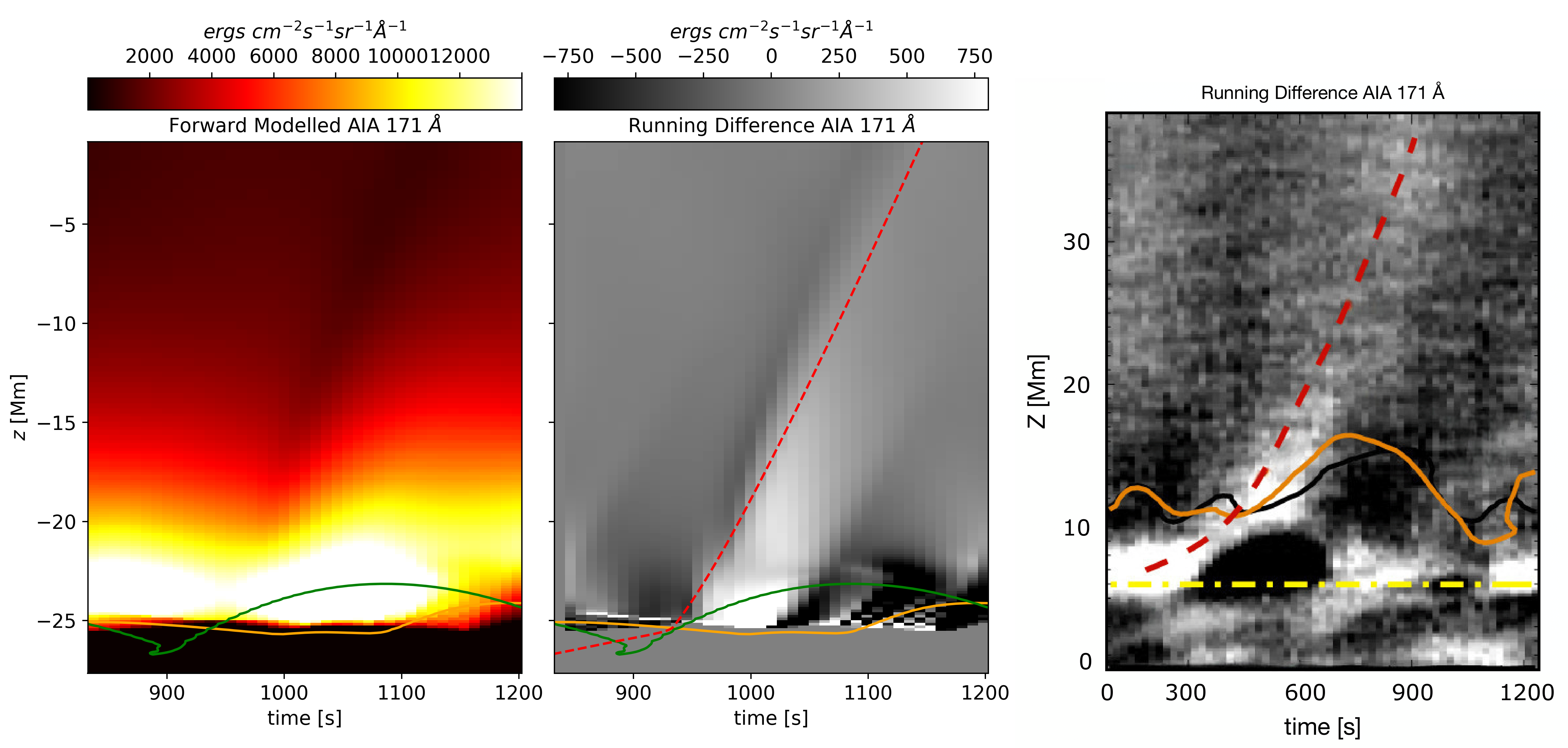}
\caption{Left: Synthetic AIA 171~\AA{} time-distance diagram produced using an artificial slit at $r=12.5$ Mm. The green contour shows the location of the transition region (T=$40,000$~K) in the simulation with a photospheric wave driver measured along field line 3, whereas the orange contour displays the same temperature contour using a vertical slit at $r = 12.5$ Mm. Middle: Running difference AIA 171~\AA{} highlighting the nature of the PD in the simulation. The red dashed line highlights the trajectory of the sound speed in the simulation domain. Right: Observation from \citet{Samanta2015} of AIA 171~\AA{} running difference where the orange and black contours outline the height of the transition region using observations from IRIS 2796~\AA{} and IRIS 1400~\AA{} channels. The red dashed line indicates the determined sound speed from the observations whereas the dashed-dot yellow line outlines the solar limb as observed in AIA instruments.}
\label{fig:fomo_analysis}
\end{figure*}

Typically, in optically thin filters, PDs appear as bright ridges in (running-difference) time-distance plots using a slit taken along the axis of a structure. Figure \ref{fig:fomo_analysis} displays the forward modelled peak intensity and running-difference time-distance diagrams for AIA 171~\AA{} from the simulation with a photospheric wave driver in comparison with the observations from \citet{Samanta2015}. For simplicity, we assumed that the angle between the structure and the observers line of sight was $0^{\circ}$, which would correspond to a structure observed off the solar limb which itself is not inclined with respect to the plane of sky. However, in the future it would be constructive to understand how projection effects may influence the observability of PDs and their relationship with the inclination of solar spicules relative to the plane of sky. We display the time-distance plots produced by taking a slice at a fixed radial position of $12.5$ Mm which corresponds to the location of field line $3$ in the corona. In Figure \ref{fig:fomo_analysis}, the chromospheric/transition region boundary can be seen as the bright region at the bottom of the plots, near $z=-21$ Mm, and the parabolic trajectory (with time) is also observed, characteristic of tracing spicular activity. In addition, we also display the height of the transition region measured from the simulations corresponding to a temperature contour of $40,000$ K. The orange contour in Figure \ref{fig:fomo_analysis} (left and middle panels) shows the height evolution of the transition region at a fixed radial slice of $12.5$ Mm, whereas the green contour shows the transition region height when measured along field line 3, which is inclined to the vertical axis, especially in the chromosphere. The left hand plot in Figure \ref{fig:fomo_analysis} shows the time-distance map of the forward modelled peak intensity in AIA 171~\AA{} and the corresponding running difference image is highlighted in the middle panel. The PDs are more easily identifiable in the running-difference image, however, they can also be located in the peak intensity time-distance plots as the edge of the dark ridge which appears. The forward modelled simulations reveal a similar conclusion to the analysis of the plasma density, in the sense that the observed PDs are seemingly launched by the transition region co-temporally with the rising of the transition region when measured along the field line. This supports the idea that slow magnetoacoustic shocks are involved in their formation as the co-temporal relationship between the transition region and the PD agrees better when the transition region height is tracked along a field line as opposed to when it is measured at a fixed spatial location. Typically, in observations of PDs, the longitudinal slit used to create time-distance plots is taken along the axis of the structure which is often inclined with respect to the vertical axis. Seen as spicules trace the magnetic field it is understandable why \citet{Samanta2015} found such a co-temporal agreement between the rising phase of spicules and the launching of PDs. The observational time-distance plot \citep{Samanta2015} is displayed in the right hand side panel of Figure \ref{fig:fomo_analysis} and there is a good agreement between the observations and the forward modelled simulations.

\begin{figure}
    \centering    \includegraphics[width=0.47\textwidth]{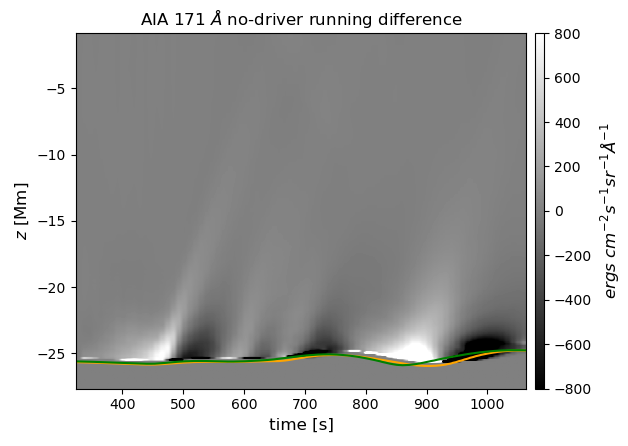}
    \caption{Running-difference AIA 171~\AA{} time-distance diagram for the simulation without a photopsheric wave driver. The green and orange lines highlight a temperature contour as outlined in Figure \ref{fig:fomo_analysis}.}
    \label{fig:additional_fomo}
\end{figure}
Figure \ref{fig:additional_fomo} shows the forward modelled AIA 171~\AA{} running-difference time-distance diagrams for the simulation without a photospheric wave driver, produced by taking an artificial slice at $r=6.5$ Mm. The forward modelled results of the simulation without a wave driver highlight the fact that faint signatures of PDs should be observable in the solar atmosphere, in associated with the dynamics of the transition region, regardless of the physical mechanisms responsible for perturbing the transition region. However, it is evident by comparing the same time-frames from both the driven and un-driven simulations that PDs appear much clearer with the presence of a wave driver, and signatures of PDs can be observed at much greater heights in the corona when photopsheric perturbations are included.

\begin{figure}
    \centering    \includegraphics[width=0.48\textwidth]{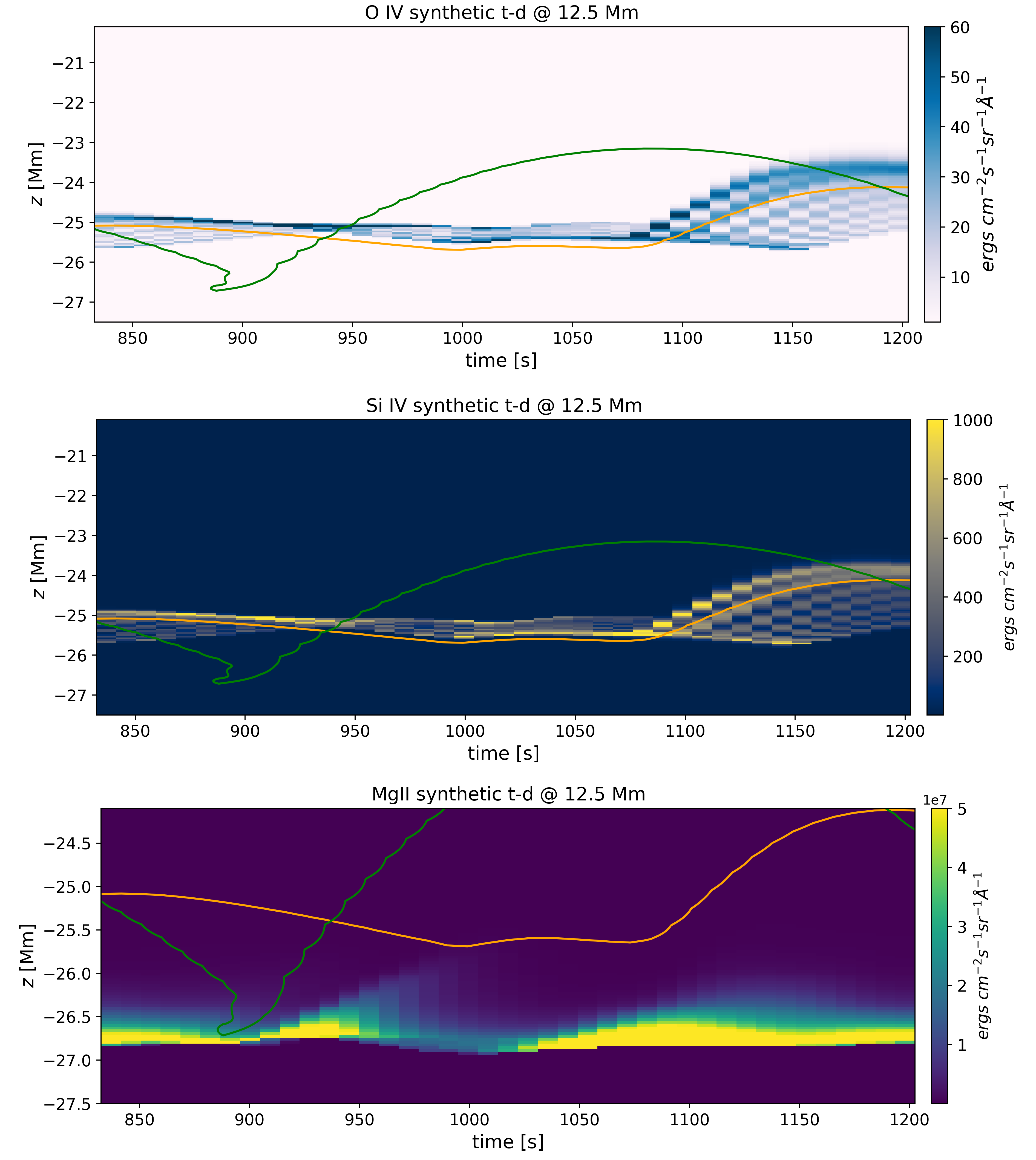}
    \caption{Time-distance plots of optically thin transition region lines O IV (top), Si IV (middle) and Mg II (bottom) from the simulation with a photospheric wave driver along a vertical slit at $r=12.5$~Mm. The height axis is zoomed in on the transition region. The green and orange lines have the same representation as outlined in Figure \ref{fig:fomo_analysis}.}
    \label{fig:fomo_MgII_withdriver}
\end{figure}
Figure \ref{fig:fomo_MgII_withdriver} displays the time-distance diagrams of the forward modelled optically thin transition region lines constructed using an artificial slice taken at $r=12.5$ Mm in the domain for the simulation with a photospheric wave driver. The chosen spectral lines (O IV, Si IV and Mg II), ranging from hottest to coolest, are some of the chromospheric spectral lines observed with IRIS \citep{DePontieu2014IRIS}. The orange and green contours in Figure \ref{fig:fomo_MgII_withdriver} represent the height of the transition region (corresponding to a temperature contour of $40,000$ K) taken at a fixed radial position of $r=12.5$ Mm and along field line 3, respectively. The forward modelled results using the hotter transition region lines (O IV and Si IV) produce a good agreement with the temperature contour of the transition region measured from the simulation using an artificial vertical slit. Therefore, if an artificial slit was chosen along the axis of the spicule formation in the simulation, then there would be a direct correlation between the transition region contour in the forward modelled analysis and the launching of a PD, as discussed in \citet{Samanta2015}. However, there is a poor agreement between the transition region height in the simulation and the forward modelled Mg II intensity. It could be argued that the forward modelling recovers the rising of the transition region on field line 3 at a time near $t=900$ s, followed by a weak signature of a propagating disturbance which appears to travel close to the sound speed in the chromosphere/transition region. However, in reality it is likely that this faint signature would be hidden amongst the stronger contribution from spicules in transition region spectral lines. In order to achieve better results regarding the forward modelling of the relationship between spicules and PDs in the chromosphere/transition region, a numerical model which accurately models the properties of the lower solar atmosphere should be employed. For example, it has been shown that by including neutrals in numerical models, spicules are amplified by magnetic tension and, through ambipolar diffusion, this can result in plasma heating \citep{mart2017}. Therefore, it is likely that the current model does not accurately reproduce the temperatures, densities and heights of solar spicules, however, their relationship and connection with PDs will not be affected. Moreover, when conducting the forward modelling, it is perhaps necessary to include optically thick radiative transfer \citep{Leenaarts2013a, Leenaarts2013b} in addition to a numerical model which incorporates the relevant physics of the chromosphere.

\section{Discussion}\label{sec:conclusions}

In this study we have numerically modelled the co-temporal relationship between the launching of propagating disturbances with the rising of the transition region, driven by chromospheric shocks, providing an explanation for their observed connection with solar spicules \citep{Jiao2015, Samanta2015}. Therefore, regardless of the physical mechanisms occurring in the lower solar atmosphere which may be responsible for driving spicules \citep[see e.g.][]{Beckers_1972, Sterling_2000, tsi2012, Skirvin2023_JET}, the formation of shocks below the transition region is sufficient to launch spicules and produce PDs with a co-temporal connection. In this study, slow magnetoacoustic waves are generated through mode conversion from p-modes at the equipartition layer \citep{Skirvin2024modeconv}, however, in the solar atmosphere slow waves may also be generated from magnetic reconnection processes \citep{McLaughlin_et_al_2012}, nonlinear coupling to Alfv\'{e}n waves via the ponderomotive force \citep{Singh2022} or thermal misbalance \citep{zaver2019,kolotkov2021}. This is important as PDs and spicules have been linked with magnetic reconnection \citep[e.g.][]{Samanta2019}, however, any mechanism producing shocks propagating along the magnetic field in the stratified solar atmosphere will be sufficient to perturb the transition region and produce PDs in the corona. In addition, there is likely a connection between the spicule jets and PDs reported in this study and the association between coronal rain and rebound shocks \citep{Fang2015, antolin2023} which should be focus of future work.

We have provided evidence for the link between PDs and solar spicules through numerical simulations both with and without the inclusion of a photopsheric wave driver, for instance, mimicking p-mode perturbations. Moreover, PDs are produced in the simulation even when a photospheric wave driver is not included. In this case, they display a periodicity which may be related to the three minute propagating coronal waves predicted from the idea of a leaky chromospheric resonator \citep{Botha2011, Snow2015, Felipe2019}. However, the presence of PDs in the corona are more clearly observed when the wave driver is included, as stronger shocks are generated in the chromosphere resulting in a greater displacement of the transition region. These propagating disturbances are intensity enhancements in the optically thin corona which travel at the local sound speed along the magnetic field, demonstrated through our forward modelling analysis of the simulations. As a result, propagating disturbances are inherently associated with properties similar to slow magnetoacoustic waves, due to their compressible nature in the low-$\beta$ corona, which transport acoustic wave energy flux along the magnetic field and may be partly responsible in the heating of local plasma. However, it should be emphasised that PDs in nature behave more like a slow magnetoacoustic pulse, their periodicity is simply a response to the periodic driver (in this case granulation in the photosphere). In any numerical model, careful attention must be paid to the physical treatment of the transition region when analysing energy flux transmission as numerical treatments of the transition region can heavily affect the calculated energy fluxes \citep{Howson2023}.
 
An interesting consequence of our analysis is that, assuming the solar atmosphere is dynamic and the transition region is constantly in motion, then PDs (and hence slow magnetoacoustic pulses) should be ubiquitous in the corona, yet slow waves/pulse are not currently observed to be omnipresent throughout the entire solar atmosphere. Investigating this discrepancy can be the focus of next generation solar telescopes such as DKIST/Solar Orbiter/MUSE. Moreover, the results presented in this work raise the question why slow waves are not observed in such large quantities in magneto-convection simulations such as MURaM \citep{vogler2005} and Bifrost \citep{Gud2011}. A possible explanation for this could be due to the high value of compressive viscosity adopted in the corona of these codes \citep{Rempel2017} which may eradicate the slow waves before they can travel a significant distance into the coronal domain.

The observational reports of \citet{Cai2024} are consistent with our results of PDs possessing periodicities apparently modulated by p-modes. The observability of PDs is more evident along pre-existing overdense structures in the optically thin solar corona, which would explain early observations of PDs from TRACE and EIT. On the other hand, PDs will inevitably be more difficult to observe in the ambient atmosphere where the density structuring is less enhanced \citep[e.g.][]{Morton_Cunningham2023}. In addition, the periodicity of the photospheric driver may influence the dynamical evolution of the transition region. In other words, a broadband wave driver may result in a transition region oscillation without a clear periodicity and investigating this should be focus for future work. 

There is also likely a connection between PDs and the mass-loading of the solar atmosphere, which is ultimately responsible for the observed inhomogeneity of the solar corona and is important in the context of driving the fast solar wind \citep{Liu2015}. There is a general consensus that solar jets resulting from magnetic reconnection are responsible for creating the fine-scale inhomogeneities in the solar corona \citep{DeForest2018, Morton_Cunningham2023,Raouafi2023}. These open-field overdense structures are a result of the mass-loading of the solar corona and may act as efficient waveguides for magnetohydrodynamic waves \citep{Banerjee2021}, with implications for wave phenomena in driving the solar wind \citep{Tian2014,Liu2015, Cranmer2017, Chitta2023}.

\begin{acknowledgements}
SJS and TVD were supported by the European Research Council (ERC) under the European Union's Horizon 2020 research and innovation programme (grant agreement No 724326) and the C1 grant TRACEspace of Internal Funds KU Leuven. TVD received financial support from the Flemish Government under the long-term structural Methusalem funding program, project SOUL: Stellar evolution in full glory, grant METH/24/012 at KU Leuven. The results received support from the FWO senior research project with number G088021N.
\end{acknowledgements}

\bibliography{ref}{}
\bibliographystyle{aa}

\end{document}